\documentclass[conference]{IEEEtran}
\IEEEoverridecommandlockouts
\usepackage{amsmath,amssymb,amsfonts}
\usepackage{algorithmic}
\usepackage{graphicx}
\usepackage{textcomp}
\usepackage{xcolor}
\usepackage{url}

\usepackage{orcidlink}

\usepackage[style=ieee, minbibnames=1, maxbibnames=1]{biblatex}
\def\BibTeX{{\rm B\kern-.05em{\sc i\kern-.025em b}\kern-.08em
    T\kern-.1667em\lower.7ex\hbox{E}\kern-.125emX}}
\begin{document}

\title{Solutions for Distributed Memory Access Mechanism on HPC Clusters\\
}

% \author{\IEEEauthorblockN{Jan Meizner}
% \IEEEauthorblockA{\textit{Sano - Centre for Computational  Medicine} \\
% Czarnowiejska 36, building C5, 30-054 Kraków,\\
% https://orcid.org/0000-0003-4094-6557}
% \and
% \IEEEauthorblockN{Maciej Malawski}
% \IEEEauthorblockA{\textit{Sano - Centre for Computational  Medicine} \\
% Czarnowiejska 36, building C5, 30-054 Kraków,\\
% https://orcid.org/0000-0001-6005-0243}
% }

\author{\IEEEauthorblockN
{
Jan Meizner\IEEEauthorrefmark{1}\orcidlink{0000-0003-4094-6557},
Maciej Malawski\IEEEauthorrefmark{1}\IEEEauthorrefmark{2}\orcidlink{0000-0001-6005-0243}
}
\IEEEauthorblockA{\IEEEauthorrefmark{1}Sano Centre for Computational Medicine, Krak\'ow, Poland\\}
\IEEEauthorblockA{\IEEEauthorrefmark{2}Faculty of Computer Science, AGH University of Krak\'ow, Poland\\}
}

\maketitle

\begin{abstract}
Paper presents and evaluates various mechanisms for remote access to memory in distributed systems based on two distinct HPC clusters. We are comparing solutions based on the shared storage and MPI (over Infiniband and Slingshot) to the local memory access. This paper also mentions medical use-cases that would mostly benefit from the described solution. We have found out that results for remote access esp. backed by MPI are similar to local memory access. 
\end{abstract}


\section{Introduction}
\label{sec:intro}
In this paper we present novel solutions for accessing memory between multiple nodes in the cluster. We aim to address multi-tenant environments such as those offered by HPC centers. Our research was prompted by various biomedical use-cases such as the Spliced Transcripts Alignment to a Reference (STAR)~\cite{Dobin2013}. In this case only around 20\% of the genome index data structure needs to be cached~\cite{Meizner2025}. This minimizes data transfer overheads over the network with higher latency than local DRAM~\cite{Shen2024}. The goal of this paper is to test various solutions for remote memory access. 

\section{Description of the Problem}
\begin{figure}[htbp]
\centerline{\includegraphics[width=\columnwidth]{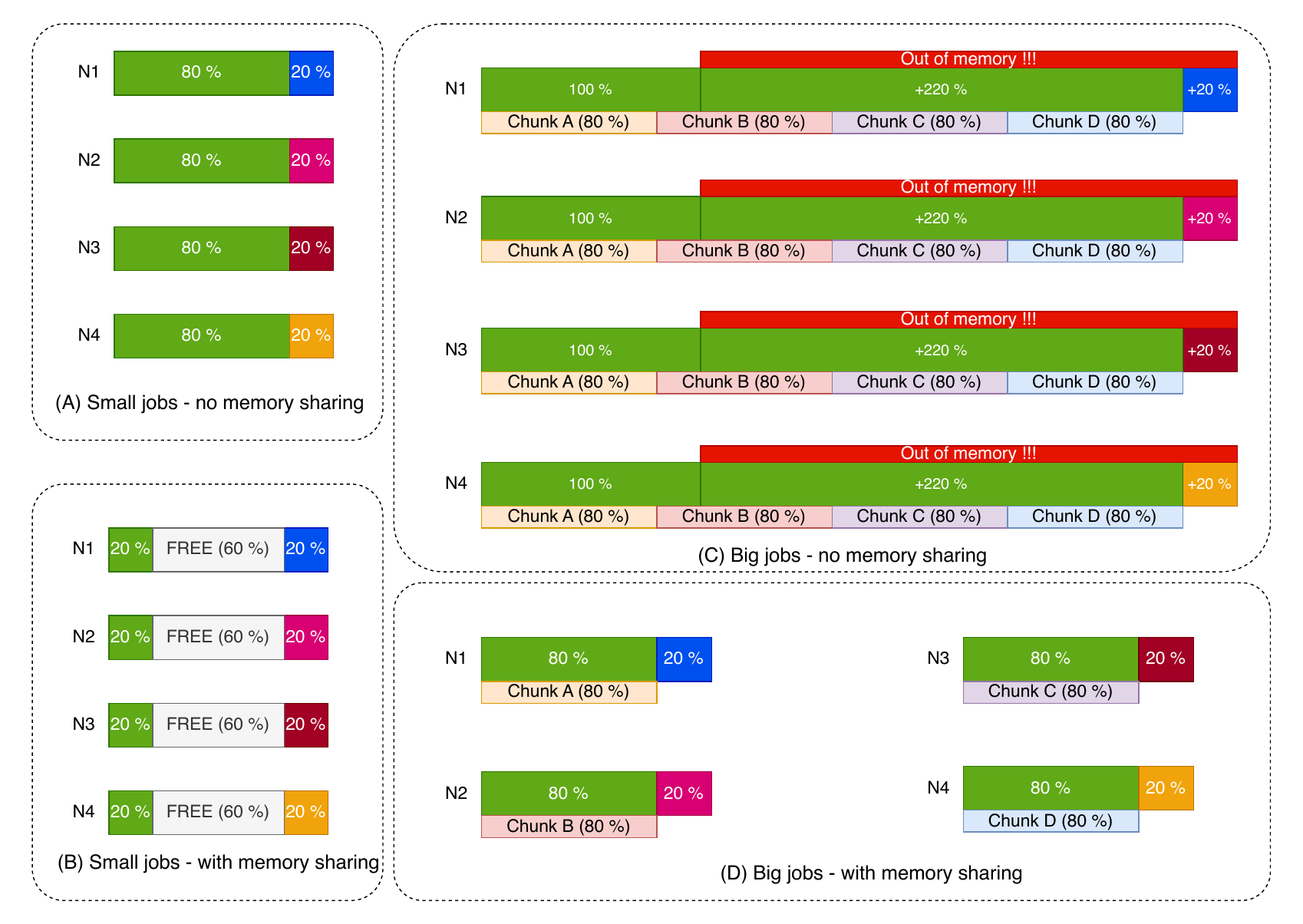}}
\caption{Sharing duplicated data (green) between nodes to conserve resources for other jobs (A, B) or to enable running bigger jobs memory-wise (C,D).}
\label{fig:problem-desc}
\end{figure}

While HPC and Cloud infrastructures offer huge computational resources it is not always sufficient as shown further in this paper. 
As mentioned in Section~\ref{sec:intro}, it is common for applications to require access to the same portion of data on multiple nodes (see Fig.~\ref{fig:problem-desc} A and B), e.g. the genome index in STAR. As shown in this diagram we may reduce total memory usage (A) by distributing portion of duplicated memory to multiple nodes (B). On the other hand there are applications that would not fit into single node or make computations cost-prohibitive due to the need to run big instances (Fig.~\ref{fig:problem-desc}C). Especially some genomics applications may require up to 1 TB of RAM. We may solve this solution by distributing chunks of duplicated data (Fig.~\ref{fig:problem-desc}D).

\section{Related Work}
\label{sec:rw}
We have found some existing solutions for the problem simmilar to ours. For example rmmap~\cite{10.1145/3627703.3629568} and "remote regions"~\cite{regions} which both rely on custom-made kernel module or modifications. There are also solutions like LegoOS~\cite{legoos} which are complete operating system designed for distributed processing. None of those solutions may be used without root-level privileges and as such are not applicable to HPC environments. In the course of our work we have also checked available libraries for inter-node communication such as the Message Passing Interface (MPI)~\cite{mpi-forum} and its implementations as well as shared storage file-systems such as Lustre~\cite{lustre}. We also analyzed low-level protocols for those solutions such as  Infiniband and o2ib based on RDMA, Slingshot and Kernel Fabric Interface (KFI) as well as LNet used by Lustre.

\section{Solution of the Problem}
\begin{figure}[htbp]
\centerline{\includegraphics[width=0.85\columnwidth]{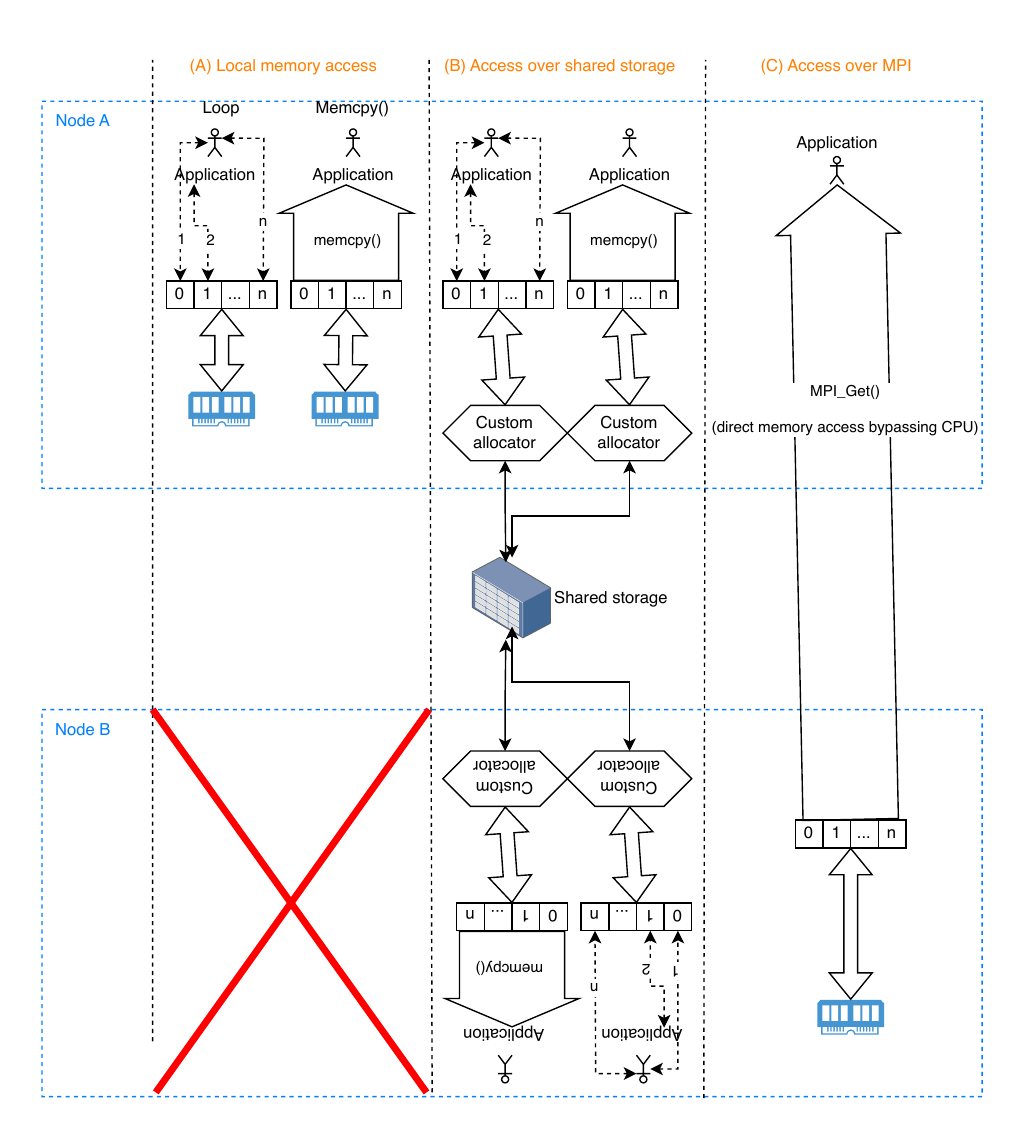}}
\caption{Proposed architectures for memory access - a baseline and two distributed.}
\label{fig:arch}
\end{figure}

Our first solution uses the LD\_PRELOAD mechanism to replace local memory access with the distributed one, keeping built-in memory allocation functions. We cannot use kernel customizations due to the lack of root-level access, so we have developed a custom library based on mmap(). It tracks memory allocations and is backed by custom-developed virtual file-system (VFS) stored on on Lustre. We have also developed a solution based on MPI which features one-way communication based purely on RDMA mechanism. We hypothesize that an MPI based solution would provide better performance than VFS. However it cannot be used with proprietary software due to the need of application code instrumentation. Those architectures are shown in Fig.~\ref{fig:arch} which also compares them with local non-distributed memory access (A) as baseline for the evaluation of our solutions. 

\section{Experiments and results}
We have performed tests on two distinct supercomputers provided by the ACC CYFRONET AGH. Ares that provides 3.5 PFlops and Helios that is the fastest supercomputer in Poland reaching over 30 PFlops. For each test we have collected 10 measurements for 10 batches of data between 100 MB and 1000 MB -- every 100 MB. For local memory and VFS we have created a piece of software in C dynamically allocating memory with malloc() function copying data with a memcpy() function. For MPI we have performed one-way data read that reflects the memory access pattern like the one used by STAR application.

\section{Discussion of the results}
\begin{figure}[htbp]
\centering
\includegraphics[width=0.22\textwidth]{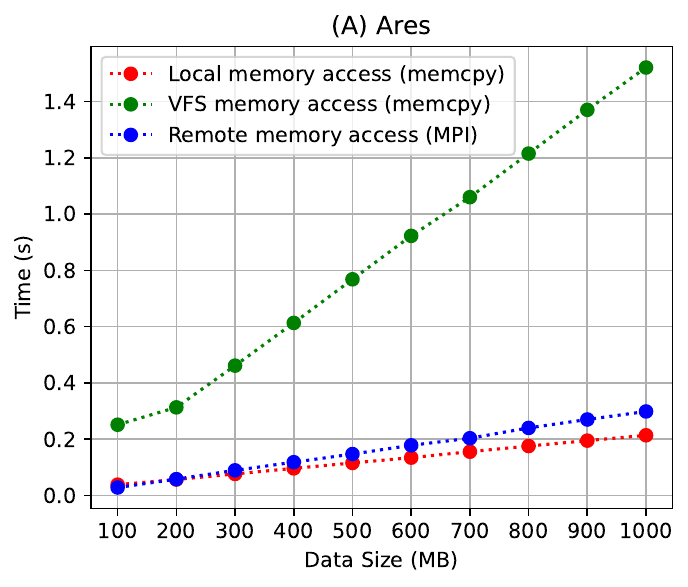}
\hfill
\includegraphics[width=0.22\textwidth]{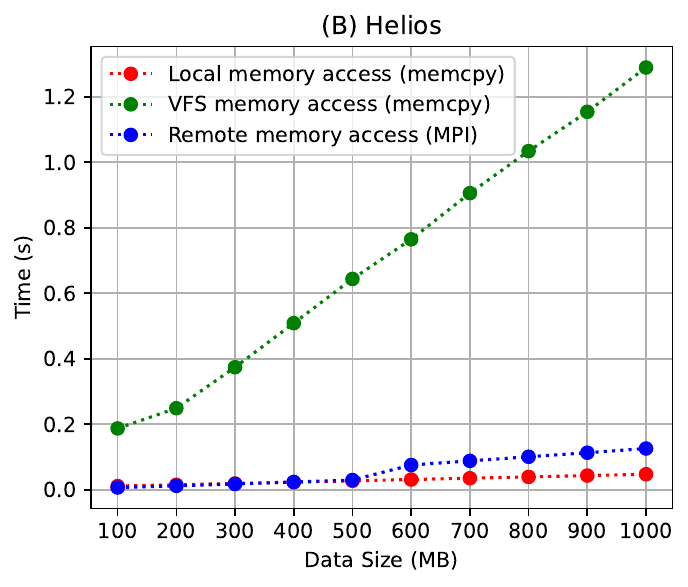}
\caption{Local memory access vs VFS vs MPI RDMA on: (A) Ares (Infiniband), (B) Helios (Slingshot).}
\label{fig:local-vs-vfs-vs-mpi}
\end{figure}

We are presenting results in Fig.~\ref{fig:local-vs-vfs-vs-mpi}. As expected the local access is faster than any remote mechanism. The VFS is prone to severe performance degradation as seen in the plot. We expect that this can be improved with faster shared storage or interconnect. On the other hand we have also found out that MPI-based memory access is comparable to local access based on memcpy() which makes it best suited for our goal. 

\section{Conclusions and future work}
We have built a solution for distributed memory access based both on VFS and the MPI. We run benchmarks on two HPC clusters. They confirmed our hypothesis that MPI-based solution provides comparable performance with the non-distributed memory access. We have also found out that VFS may be too slow for pessimistic use-case but still can be used for moderately short jobs requiring huge amount of memory -- quite common for example in genomics. To better address  closed-source applications we are planning to create a custom solution based on File System in Userspace (FUSE) and MPI.

\section*{Acknowledgment}

This publication is partly supported by the EU H2020 grants Sano (857533), NEARDATA (101092644) and by the Minister of Science and Higher Education "Support for the activity of Centers of Excellence established in Poland under Horizon 2020" number MEiN/2023/DIR/3796. We gratefully acknowledge Polish high-performance computing infrastructure PLGrid (HPC Center: ACK Cyfronet AGH) for providing computer facilities and support within computational grant no. PLG/2025/018235.

% \bibliographystyle{IEEEtran}
% \bibliography{main}

\printbibliography

\end{document}